\documentclass[11pt]{article}

\usepackage[latin1]{inputenc}
\usepackage{def,epsf}
\evensidemargin\oddsidemargin
\input{psfig.sty}

\begin{document}

\title
      {Photon Stars}

\author{
        Heinz-J\"urgen Schmidt,
        Felix Homann\\
        Fachbereich Physik,
        Universit\"at Osnabr\"uck,
        Postfach 4469, \\
        D-49069 Osnabr\"uck, Germany}

\date{M\"arz 1999 }

\maketitle

\begin{abstract}
We discuss numerical solutions of {\sc Einstein}'s field equation describing static, spherically
 symmetric conglomerations of a photon gas. These equations imply a back reaction of the metric
 on the energy density of the photon gas according to {\sc Tolman}'s equation. The 3-fold of solutions
corresponds to a class of physically different solutions which is
parameterized by only two quantities, e.~g.~mass and surface
temperature. The energy density is typically concentrated on a shell
because the center contains a repelling singularity, which can, however,
not be reached by timelike or null geodesics. The physical
relevance of these solutions is completely open, although their
existence may raise some doubts w.r.~to the stability of black
holes.
\end{abstract}

\section{Introduction}

The starting point of this investigation was the discussion of the
Carnot-Bekenstein-process in the environment of a {\sc Schwarzschild}
black hole \cite{HUS}. There it was assumed that the black hole is
surrounded by a cloud of radiation with a local temperature
according to {\sc Tolman}'s equation \cite{TOL}
        \begin{equation}
                T(r)= \frac{T_\infty}{ g_{00} (r)}     \label{1}
        \end{equation}
where $ T_{\infty} $ is the usual {\sc Hawking} temperature at $ r=\infty $. This equation also follows
for the equilibrium distribution of photons in relativistic kinetic theory \cite{NEU} or from
elementary thermodynamical gedanken experiments \cite{HUS}. These yield a modified formula
for the efficiency  of the Carnot-Bekenstein-process
        \begin{equation}
\eta = 1- \frac{ T_2 \sqrt{ g_{00}(2) } }{ T_1 \sqrt{ g_{00}(1) }}  \label{2}
        \end{equation}
between two heat reservoirs at different height. In the equilibrium we have $ \eta = 0 $ which
implies (\ref{1}). All this holds as long as the back reaction of the radiation to the metric can
be neglected.

But, since $ g_{00} (r) = 1 - \frac{ {\cal R} }{r} $, at the
{\sc Schwarzschild} radius $r= {\cal R} $ the temperature and the energy
density diverges and cannot longer be regarded as a mere
perturbation. Rather one would have to solve the field equation
anew, this time allowing for the energy-stress tensor of the photon
gas to act as a source of gravitation. It is not at all clear
whether the resulting metric will be  a modified black hole in some
sense. So we have the following problem which may be considered
independently of the original motivation:

Calculate the static, spherically symmetric metric of {\sc Einstein}'s field equations with the
energy-stress tensor of a perfect fluid
        \begin{equation}
T_{ab} = ( \rho + P ) u_a u_b + P g_{ab}                \label{3}
        \end{equation}
consisting of photons, i.e.
        \begin{equation}
\rho = 3P.                                      \label{4}
        \end{equation}

In section 2 the corresponding field equations are transformed into an autonomous
two--dimensional system of differential equations which is discussed in section 3.
In section 4 we investigate the metric of a photon star for $r\to 0$ and study its global properties
by means of numerical solutions. Physical characteristics like radius, mass and temperature
are defined in section 5. Section 6 contains concluding remarks.

\section{Transformation of the field equations}

With except of eq.~(\ref{4}) the problem stated above is just the
well-known problem of constructing the interior solution of a star.
We may choose coordinates $(t,r,\theta,\phi)$ such that the metric
is given by
        \begin{equation}
ds^2= - f(r) dt^2 + h(r) dr^2 + r^2 d\Omega^2   \label{5}
        \end{equation}
where $f$ and $h$ are unknown functions and $ d\Omega^2 $ is the
surface element of a unit sphere. With respect to these coordinates
the field equations boil down to a system of three coupled
differential equations ( cf. \cite{WAL} 6.2.3--6.2.5)
        \begin{eqnarray}
\frac{ 8\pi G}{c^2} \rho &=& \frac{ h'}{rh^2} + \frac{h-1}{hr^2}                \label{6}       \\
\frac{ 8\pi G}{3c^2} \rho &=& \frac{ f'}{rfh} - \frac{h-1}{hr^2}                \label{7}       \\
\frac{ 16\pi G}{3c^2} \rho &=& \frac{f'}{rfh}- \frac{h'}{rh^2} + \frac{1}{\sqrt{fh}}
                \frac{d}{dr} \left( \frac{f'}{\sqrt{fh}} \right)        \label{8}
        \end{eqnarray}
We recall that {\sc Tolman}'s equation \cite{TOL} was derived under the
same assumptions we made, except spherical symmetry. Hence we may
adopt (\ref{1}) or, equivalently,
        \begin{equation}
\rho(r)= \frac{\rho_1}{f^2(r)} ,                                                \label{9}
        \end{equation}
since
        \begin{equation}
\rho = \sigma T^4                                                       \label{10}
        \end{equation}
from local statistical mechanics, where $ \sigma$ is the
Stefan-Boltzmann constant. We write
        \begin{equation}
\rho_1 = \frac{c^4}{8 \pi G} C                                          \label{11}
        \end{equation}
and obtain from (\ref{6}) and (\ref{7})
        \begin{eqnarray}
\frac{C}{f^2}  &=& \frac{h'}{r h^2} + \frac{h-1}{h r^2}                 \label{12}  \\
\frac{C}{3 f^2} &=& \frac{ f'}{ r f h } - \frac{h-1}{ h r^2} .          \label{13}
        \end{eqnarray}
Eq.~(\ref{8}) is identically satisfied, which confirms {\sc Tolman}'s
result. Otherwise the system of differential equations would be
overdetermined.

The set of solutions ${\cal S}$ of the system (\ref{12}),
(\ref{13}) can be parameterized by 3 parameters, e.~g.~$C$ and two
initial values for $f$ and $h$. But $ {\cal S}$ is invariant under
the 2-parameter group of scale transformations
        \begin{equation}
r \mapsto \lambda r,\quad  f \mapsto \mu f, \quad h \mapsto h,
\quad C
\mapsto
\frac{\lambda^2}{\mu^2} C
                                                                        \label{14}
        \end{equation}
Thus there exists only a 1-parameter family of solutions looking
qualitatively different.

\noindent An equivalent second order equation is obtained by eliminating $h$
using
        \begin{equation}
h = \frac{ 3 f (f+r f')}{C r^2 + 3f^2}                                  \label{15}
        \end{equation}
and inserting the derivative of (\ref{15}) into (\ref{12}):
        \begin{equation}
f''=\frac{6rf^2 + 6r^2 f f'- 6 f^3 f' + 2r^3 f'^2}{ C r^3 f + 3 rf^3}   \label{16}
        \end{equation}

\noindent For $ C=0$ we have the well-known equations which lead to a 2-parameter family of {\sc Schwarzschild}
metrics. For $ C \neq 0$ we may scale every solution such that it becomes a solution with
        \begin{equation}
C=1.                                                                     \label{17}
        \end{equation}
The remaining subgroup of (\ref{14}) with $\lambda = \mu$ may be
used to simplify the differential equation (\ref{16}). We perform
the transformation
        \begin{equation}
s= \ln \frac{r}{r_0}, \quad x= \frac{f(r)}{r}, \quad y= f'(r) + x,
\label{18}
        \end{equation}
where $ r_0 > 0$ is arbitrary. The resulting reduced equations read
        \begin{eqnarray}
\frac{dx}{ds} &=&  y-2x                                                 \label{19}      \\
\frac{dy}{ds} &=& \frac{ y (2y-3x (x^2-1)) }{ x(3x^2+1) }.              \label{20}
        \end{eqnarray}
Note that this system is autonomous. The resulting symmetry $ s
\mapsto s+s_0 $ reflects the scale invariance of (\ref{16}) w.r. to
the subgroup $\lambda = \mu$. Even if we cannot solve it exactly,
it seems to be better accessible for intuition and developing
approximation schemes. Any two different solutions of (\ref{19}),
(\ref{20}) correspond to different similarity classes of ${\cal
S}$.

\section{Discussion of the reduced equations}

We calculate some typical solutions of (\ref{19}), (\ref{20}) numerically and display them as
curves in the $x-y$-plane.

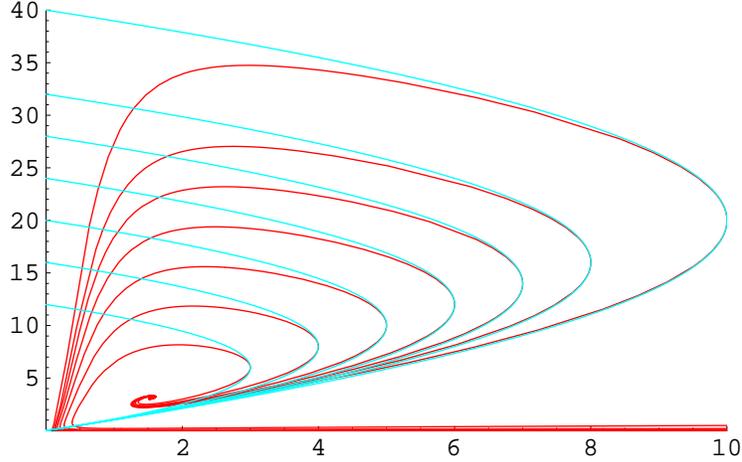
\FIGU{dgl2}{100mm}{Figure 1: A selection of numerical solution curves of the reduced equation (\ref{19}), (\ref{20})  
                 together with the parabolic {\sc Schwarzschild} approximation} 

\noindent It is obvious that
        \begin{equation}
x_0 = \sqrt{ \frac{3}{7} }, \quad   y_0 = 2 \sqrt{ \frac{3}{7} }
\label{21}
        \end{equation}
is a stable stationary point of (\ref{19}), (\ref{20}) which is an
attractor of the whole open quadrant $ x>0,  y>0$. It follows by
inverting the transformation (\ref{18}) that
        \begin{equation}
f(r)= \sqrt{ \frac{7}{3} } r , \quad  h(r) = \frac{7}{4}
\label{22}
        \end{equation}
is an exact solution of (\ref{19}), (\ref{20}) which is
asymptotically approached by any other solution for $ r \to \infty
$. It follows that the space-time of photon stars is not
 asymptotically flat. The other exact solution of (\ref{19}), (\ref{20})
        \begin{equation}
y=0, \quad   x= a e^{-2s}               \label{23}
        \end{equation}
yields
        \begin{equation}
f(r)= \frac{a}{r}, \quad   h(r) = 0     \label{24}
        \end{equation}
and is hence unphysical. A typical solution curve of (\ref{19}),
(\ref{20}) starts from $ x=+ \infty $, $y=0$, $ s=- \infty$ and
runs close to the solution (\ref{23}) until it reaches small values
of $x$. Then according to the "$x$" in the denominator of
(\ref{20}), $\frac{dy}{ds} $ increases rapidly and the curve is
turned up towards the $y$-axis. Then it describes a parabolic-like
bow and approaches the stationary point by a clockwise vortex.

It is instructive to draw the general {\sc Schwarzschild} solutions (
$C=0$)
        \begin{equation}
f_S(r)=a- \frac{b}{r}                                           \label{25}
        \end{equation}
into the $x-y$-diagram. They are given by the family of parabolas
        \begin{equation}
x=y-\frac{b}{a^2} y^2,                                          \label{26}
        \end{equation}
which approximate the solution curves of (\ref{19}), (\ref{20})
having the same vertex at
        \begin{equation}
x_1= \frac{a^2}{4b}, \quad   y_1 = \frac{a^2}{2b}.
\label{27}
        \end{equation}
In this way, for each solution $ \left< f(r), h(r) \right> $ of
(\ref{12}), (\ref{13}) we can define a unique {\sc Schwarzschild}
approximation $ \left< f_S(r), h_S(r) \right>$.

\section{Properties of the metric}

We now turn to the discussion of the solutions of (\ref{16}) for $f(r)$ which yield $h(r)$ by
(\ref{15}).

\noindent To study the behaviour for smaller $r$ we expand $f$ into a Laurent
series and insert this series into (\ref{16}). It turns out that
the series starts with
        \begin{equation}
f(r)=\frac{A}{r} + B + \cdots                                   \label{28}
        \end{equation}
in accordance with a {\sc Schwarzschild} solution for $C=0$.
The coefficients $A,B$ are left undetermined since they represent
 the two initial values for (\ref{16}). The next terms are uniquely determined by (\ref{16}).
It is straight forward to calculate the first, say, 20 terms by using a computer algebra
software like MATHEMATICA. Here we only note down the first nonvanishing extra terms for $f$
and $h^{-1}$:
        \begin{eqnarray}
f(r) &=& \frac{A}{r}+ B + \frac{CB}{15 A^2} + {\cal O}(r^5),    \label{29}      \\
h^{-1}(r) &=& \frac{A}{Br} +1 - \frac{C}{15 A^2} r^4 + {\cal O}(r^5).   \label{30}
        \end{eqnarray}
For small $r$, $f$ and $h$ are also approximated by {\sc Schwarzschild}
solutions, but unlike the approximation discussed above, we have to
choose $A>0$ in order to obtain a positive solution $f(r)>0$. $f$
cannot change its sign in a continuous way, since $f(r_0)=0$
implies $f''(r_0) = \infty$ by (\ref{16}). (According to the scale
invariance (\ref{14}), $f(r)$ may be multiplied by $-1$, but this
gives no physically different solution.) Thus we may state that for
$r \to 0$ the metric looks like that of a {\sc Schwarzschild} black hole
with negative mass, independent of $C$. For the geodesic motion
close to $r=0$ we may thus adopt the effective potential of
{\sc Schwarzschild} theory ( \cite{WAL} 6.3.15) (with $M \mapsto -M$):
        \begin{equation}
V=\frac{1}{2} \kappa + \kappa \frac{M}{r} + \frac{L^2}{2r^2} + \frac{ML^2}{r^3} \label{31}
        \end{equation}
where
        \begin{equation}
\kappa = \left\{ \begin{array}{r@{\quad}l}
                        1 & \mbox{ (timelike geodesics)} \\
                        0 & \mbox{ (null geodesics)}
                \end{array}
        \right.                                                 \label{32}
        \end{equation}
It follows that $r=0$ can never be reached by particles or photons
due to the infinite high potential barrier. Although curvature
blows up for $r \to 0$, as in the {\sc Schwarzschild} case, the nature of
the singularity is less harmful. We conjecture that it
cannot be regarded as a "naked singularity" in whatever technical
sense (see \cite{EAR} for details of the various
 definitions) and that Cauchy surfaces still exist in the spacetime given by (\ref{12}),
(\ref{13}), if the line $r=0$ is excluded from the spacetime manifold.

\noindent From a computational point of view, the singularity of $f(r)$ at
$r=0$ suggests to transform (\ref{16}) into a differential equation
for
        \begin{equation}
F(r):=r f(r)                                                    \label{33}
        \end{equation}
and to re-transform to $f(r)$ after a numerical solution for $F$
has been obtained. We used the {\tt NDSolve}-command of MATHEMATICA
to produce the following numerical solutions:

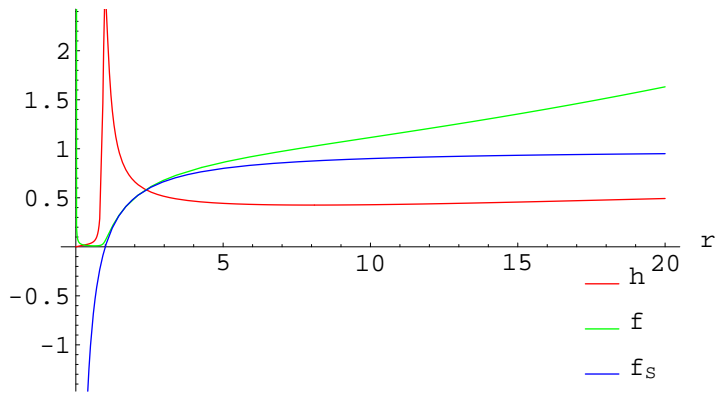
\FIGU{fhs}{100mm}{Figure 2: A typical numerical solution $f(r), h(r)$ of the system
                 (\ref{12}), (\ref{13}). $h(r)$ has its maximum at $r=r_0$.
                  The corresponding {\sc {\sc Schwarzschild}} solution
                 $f_S(r)$ with  $f_S(r_0)=0$  is also displayed.}    

A typical solution is shown in Fig.~2. Recall that   for the {\sc Schwarzschild}
metric $h(r)$ diverges at $r={\cal R}$ and $f({\cal R})=0$. For the solution of Fig.~2 $h(r)$
has a relatively sharp maximum at $r_0$ and $f(r_0)$ is becoming small. For $r>r_0$, $f$ and
 $h$ are comparable with their {\sc Schwarzschild} approximations $f_S$ and $h_S$, if $r$ is not too
large. For $r<r_0$, $f$ remains small within some shell $ r_1<r<r_0 $ and diverges for
$ r \to 0$ according to (\ref{29}). By (\ref{9}) this means that the energy density is
concentrated within that shell and $r_0$ may be viewed as the "radius of the photon star".

Other solutions with larger values of $C$ show a more diffuse cloud of photons and a less
sharp maximum of $h$, see Fig.~3 and 4. These solutions are all scaled to the same value of
$r_0$.

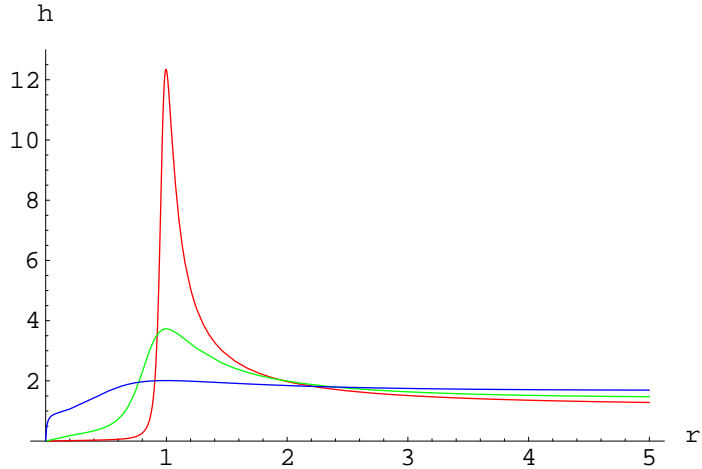
\FIGU{v_h}{100mm}{Figure 3: Numerical solutions $h(r)$ for different $C$. The solutions are scaled 
such that they obtain their maximum at the same value $r_0=1$.}

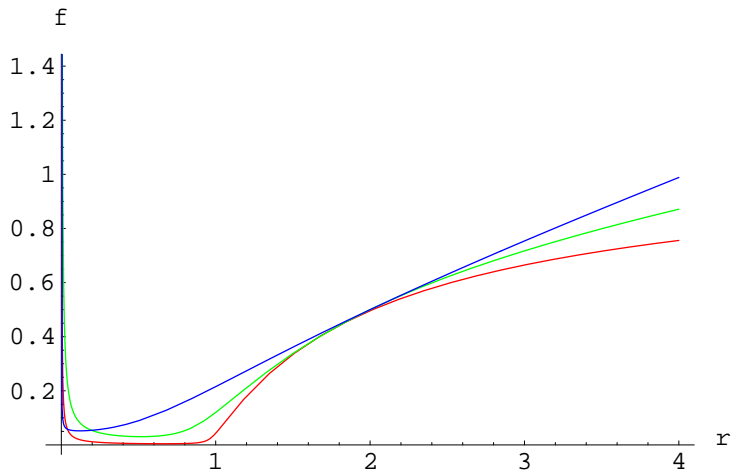
\FIGU{v_f}{100mm}{Figure 4: Numerical solutions $f(r)$ for different $C$ and the same scaling as in Fig.~3.}

\section{Physical parameters of a photon star}
We have seen that the set of solutions ${\cal S}$ may be
characterized by 3 parameters, e.~g.~$A,B$ and $C$ in (\ref{29}).
From the analogy with the {\sc Schwarzschild} case ($C=0$) we expect that
only a 2-parameter family represents physically different
spacetimes. In the {\sc Schwarzschild} case, one parameter is set to 1 by
the choice of the units, and the remaining parameter ${\cal R}$
distinguishes between black holes of different mass. More
specifically, one postulates that the velocity of light, expressed
by $\frac{dr}{dt}$ approaches 1 for $r \to \infty$. The metric then
obtains the form
        \begin{equation}
f_S(r)=1-\frac{\cal R}{r}, \quad h_S=f_S^{-1}.                          \label{34}
        \end{equation}
In the case of the photon star, we cannot proceed in the same way,
since the metric will not be asymptotically flat. But instead we
may postulate the "gauge condition" that the {\sc Schwarzschild}
 approximation of $ \left< f, h \right> $, defined in section 3, should obey condition
(\ref{34}). If this is not the case, one has to perform a suitable
scale transformation (\ref{14}). In this way we obtain a two-fold
of physically different solutions.

\noindent We now consider physical parameters characterizing this two-fold of
solutions. One could be the radius $r_0$ of the photon star defined
above.

\noindent In analogy to the {\sc Schwarzschild} theory (\cite{WAL} 6.2.7) we introduce the (gravitational)
mass function
        \begin{equation}
m(r):=\frac{r}{2} (1-h^{-1}(r) )\frac{c^2}{G}.                          \label{35}
        \end{equation}
In the domain where $ f(r) \approx f_S(r)$ this is the
"would-be-mass" of an equivalent black hole. In the domain $ r<r_0$
the interpretation of (\ref{35}) is not so obvious. As to be
expected from the above discussion of the metric for $r \to 0$, it
turns out that $m(0)<0$. A typical mass function is shown in Fig.~5, where also the "proper mass"
        \begin{equation}
m_p(r) = \frac{4\pi}{c^2} \int_0^r \rho(r')h(r')^{\frac{1}{2}} dr'      \label{36}
        \end{equation}
and the difference $m_p - m$ is displayed.

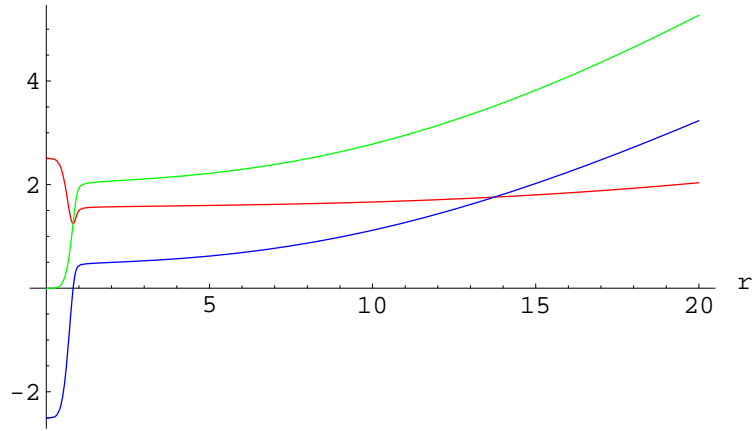
\FIGU{m}{100mm}{Figure 5: A typical numerical solution of the mass functions $m(r), m_p(r)$ and the difference
                $m_p(r) - m(r)$.}

\noindent It may be, as in this case, that the majority of the photons are "hidden" by the apparent
negative mass in the center with respect to gravitation.

\noindent Nevertheless, we could use
        \begin{equation}
m_0=m(r_0)                                                              \label{37}
        \end{equation}
as a further physical parameter characterizing a photon star.
Generally, by (\ref{35}) and
$h(r_0)>\frac{4}{7}=\lim_{r\rightarrow\infty}h(r)$
        \begin{equation}
\frac{3}{7}M_S(r_0)<m_0<M_S(r_0),                                       \label{38}
        \end{equation}
where
        \begin{equation}
M_S(r_0):=\frac{r_0 c^2}{2G}.                                           \label{38a}
        \end{equation}
If $h(r_0)\gg 1$ we have $m_0\approx M_S(r_0)$, as in the {\sc Schwarzschild} case.\\

\noindent As another physical parameter we consider the "surface temperature"
        \begin{equation}
T_0:=T(r_0)=\left(\frac{\rho(r_0)}{\sigma}\right)^{1/4} .               \label{39}
        \end{equation}

\noindent If we have only a two-fold of physically different solutions, as we claimed above,
$T_0$ should be a function of $r_0$ and $m_0$. Indeed, (\ref{12}) together with
$h'(r_0)=0$ shows that $f(r_0)$ is a function of $r_0$ and $h(r_0)$, hence
of $r_0$ and $m_0$. Then, by (\ref{39}) and (\ref{9}), also $T_0$ depends only on
$r_0$ and $m_0$. Since $\sigma$ depends on $\hbar$, the result can be conveniently
expressed by using {\sc Planck} units, indicated by a subscript P:

        \begin{equation}
\frac{T_0}{T_P}=\frac{15^{1/4}}{(2\pi)^{3/4}}
\left[\frac{m_0}{M_P}\left( \frac{L_P}{r_0} \right)^3 \right]^{1/4}.    \label{40}
        \end{equation}

To give a numerical example, if we take $m_0$ as the mass of the sun,
$r_0$ as the corresponding {\sc Schwarzschild} radius, we obtain $T_0\approx 4\cdot 10^{12}$K.
This would correspond to a very hard gamma radiation with a wavelength
$\lambda \approx 10^{-15}$ m. The {\sc {\sc Hawking}} temperature of this example is
$T_H\approx 10^{-8}$K, since $T_H\sim m^{-1}$ whereas $T_0\sim m_0^{-1/2}$ for
$m\approx M_S(r_0)$.

\section{Conclusion}

It is difficult to assess the physical relevance of our findings.
But one point seems to be clear: The global character of the
solutions $\langle f,h \rangle$ is completely different from the
{\sc Schwarzschild} approximations $\langle f_S,h_S \rangle$, no matter
how small $C$ is. So the class ${\cal S}$ of solutions of
(\ref{12}), (\ref{13}) does not depend continuously on $C$ in any
reasonable sense. Any small amount of radiation will destroy the
event horizon, at least if an equilibrium is approached. If this
would also be the case in simulations of the birth of
"black holes" by collapsing matter, then they would never be born,
and perhaps won't exist at all.

\end{document}